\documentclass[10pt,letterpaper,twocolumn]{article} 
\usepackage{ol2}
\usepackage[draft,implicit=false]{hyperref}
\usepackage{amsmath}

\begin{document}

\twocolumn[ 

\title{\emph{Cavity controlled spectral singularity}}

\author{K. Nireekshan Reddy and S. Dutta Gupta$^{*}$ }

\address{
School of Physics, University of Hyderabad, Hyderabad-500046, India
\\
$^*$Corresponding author: sdghyderabad@gmail.com
}

\begin{abstract}
We study theoretically a $\mathcal{PT}$-symmetric saturable balanced gain-loss system in a ring cavity configuration. The saturable gain and loss are modeled by two-level medium with or without population inversion. We show that the specifics of the spectral singularity can be fully controlled by the cavity and the atomic detuning parameters. The theory is based on the mean-field approximation as in standard theory of optical bistability. Further, in the linear regime we demonstrate the regularization of the singularity in detuned systems, while larger input power levels are shown to be adequate to limit the infinite growth in absence of detunings.
\end{abstract}


 ] 
In recent years there has been a great deal of interest in spectral singularities in non-Hermitian systems, especially in $\mathcal{PT}$(parity-time) symmetric systems \cite{muga,ali-prl-09,zafar-ss,ali-th,ali-wg,longhi-ss}. Such singularities with diverging scattered amplitudes (in reflection and transmission) were initially studied  because of fundamental interest. A great deal of research has been devoted to understand the origin and the nature of these singularities and also the mechanism to regularize the infinities, since real systems cannot support such divergence. Indeed, no realistic system can ever support the infinite growth and such growing amplitudes will eventually render the response of the medium nonlinear. It was further understood that a dispersive Kerr nonlinearity is inadequate to regularize the singularity \cite{ali-kerr}. It was shown recently that regularization can be achieved by an all-order saturation mechanism in a balanced gain-loss system \cite{liu}. In addition to the fundamental interest in $\mathcal{PT}$-symmetric systems, there have been few attempts to find some basic applications in photonics \cite{ap1,ap2,ap3,ap4,ap5,ap6}. But any application would require a thorough control of the singularities. In this paper we show that such a control can be achieved if the gain and the loss media are kept in separate cavities with an overall feed-back provided by a third cavity. A similar ring cavity configuration with saturable absorbers was studied earlier in the context of demonstrating remarkable flexibility in the multi-stable response \cite{gsa}. Here we exploit the same flexibility albeit with a replacement of the saturable absorber with a saturable gain medium in one of the cavities. The balanced loss and gain cavities with identical parameters form the core of the $\mathcal{PT}$-symmetric system. We first investigate the linear response which clearly shows the singularity for a perfectly tuned system as reported by others \cite{ali-prl-09}. We show how $\mathcal{PT}$-symmetry is violated and these are regularized by introduction of the atomic and the cavity detunings leading to poles in scattered amplitudes away from the real axis. We then look at the nonlinear response which reiterates the earlier findings that saturable nonlinearity is able to limit the infinite growth. We further report on the bistable response and the non-reciprocity in transmission \cite{liu} which can lead to switching and optical isolation or diode action \cite{isolator}.
\begin{center}
\begin{figure}[h]
\includegraphics[width=8cm]{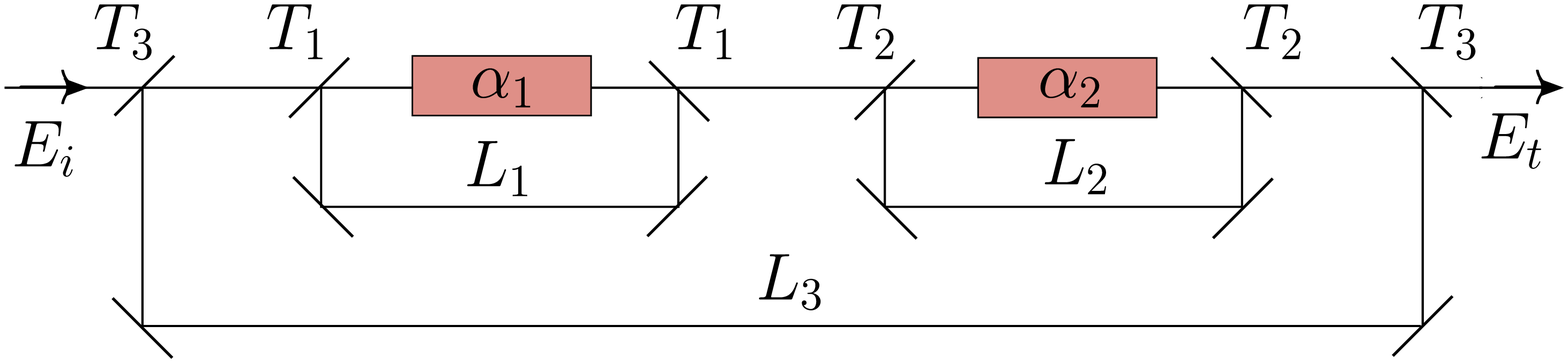}
\caption{Schematic view of the $\mathcal{PT}$-symmetric ring cavity system with coupled cavities with balanced gain and loss medium.}%
\label{fig1}
\end{figure}
\end{center}
Consider the system shown in Fig.~\ref{fig1} comprising of two coupled ring cavities of length $L_1$ and $L_2$ and mirror transmission $T_1$ and $T_2$, respectively, enclosed in a third cavity  with mirror transmission given by $T_3$. All the cavity mirrors are assumed to be lossless (i.e.,$T+R=1$, $R$- reflection coefficient). As mentioned earlier, optical multi-stability was investigated in a similar system with both the cavities containing lossy media \cite{gsa}. Unlike the general cavity configuration as in Fig.~\ref{fig1}, we consider a special case comprising of two identical coupled ring cavities of equal length ($L_1=L_2=L$) and mirror transmission ($T_1=T_2=T$). In order to realize $\mathcal{PT}$ symmetry, we further assume that the internal cavities are balanced with the same gain ($\alpha_1<0$) and loss ($\alpha_2>0$) coefficient ie., $|\alpha_1|=\alpha_2=\alpha$. This symmetry additionally implies that $\delta_1=\delta_2=\delta$ and $\theta_1=\theta_2=\theta$, where $\delta$ and $\theta$ give the atomic and the cavity detunings, respectively. Note that the exterior cavity can also be detuned from its resonance leading to a finite $\theta_3$. For such a case the input $y$ vs. output $x$ relation is given by
\begin{eqnarray}
y & = & x\left \lbrace \vphantom{\frac{ [ \tau + \zeta \sqrt{T_3/T} ] } { 1 + \delta^2 + \left| x [ \tau + \zeta \sqrt{T_3/T} ] \right|^2 } } \sqrt{\frac{T}{T_3}} \left[ \tau^2 -R_3 (1-i T_3 \theta_3) \right]+ \zeta \tau \right. \nonumber \\ \label{eq1} 
& - & \left.\frac{\zeta  (1+\delta^2+|x|^2)\left[ \tau + \zeta \sqrt{T_3/T} \right] } { 1 + \delta^2 + \left| x \left[ \tau + \zeta \sqrt{T_3/T} \right] \right|^2 } \right \rbrace, \\ \hspace{-2cm} \text{with} \nonumber \\ 
\zeta & = &\frac{\alpha L (1-i\delta)}{\sqrt{TT_3} (1+\delta^2+|x|^2)},~~~ \tau=1+iR\theta \label{eq2}.
\end{eqnarray}
The parameters in Eq.(\ref{eq1}) are given by \cite{lugiato}
\begin{eqnarray}
\label{eq3}
x  & = & \frac{\mu_2}{\hbar} \frac{E_t}{\sqrt{TT_3\gamma_{\|}\gamma_{\bot}}}, ~~~~~~~~y= \frac{\mu_2}{\hbar} \frac{E_i}{\sqrt{\gamma_{\|}\gamma_{\bot}}} ,\\ \label{eq4}
\delta_i & = & \omega_i-\omega_0 ,~~\theta_i=(\omega_{ci}-\omega_0)/\kappa, ~~ \kappa=cT/L.
\end{eqnarray}
$\omega_i$ is the atomic transition frequency in the $i^{th}$ cavity, $\omega_0$ and $\omega_{ci}$ are the incident light and the $i^{th}$ cavity frequency. $R_i$ and $T_i$ give the intensity reflection and transmission coefficients of the lossless mirrors of the $i^{th}$ cavity. It is clear from Eq.~(\ref{eq1}) that in the linear regime ($|x|\rightarrow 0$) the occurrence of the the spectral singularity requires the vanishing of the expression in the curly brackets. This implies, for example, for perfectly tuned system the following simple relation for $\alpha L$ 
\begin{equation}
\label{eq5}
\alpha L=\sqrt{T_3T^2}.
\end{equation}
It is evident from Eq.~(\ref{eq5}) that the better the cavities the easier it is to obtain the spectral singularity at lower gain levels or smaller systems. In what follows we present a detailed study of the linear response of the system and show the occurrence of the singularity and its dependence on the cavity parameters.
We consider three specific cases:
\subsection*{Effect of the exterior cavity, no internal cavities}
\begin{figure}[h]
\center{\includegraphics[width=8cm]{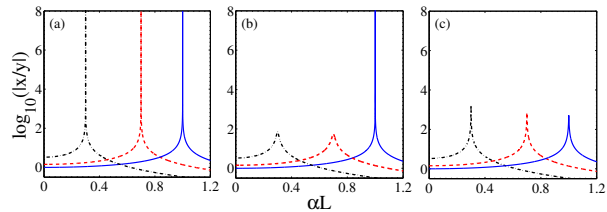}}
\caption{Effect of the external cavity on the spectral singularity in absence of internal cavities ($T=1.0)$: (a) $\delta=0.0$, $\theta_3=0.0$ (b) $\delta=0.0$, $\theta_3=0.05$, (c) $\delta=0.001$, $\theta_3=0$. Curves from right to left are for $T_3=1.0,~0.49,~0.09$.}
\label{fig2}
\end{figure}

We assume the mirrors of the internal cavities to be completely transparent ($T=1$) and we look at the influence of the exterior cavity. The results for $\log_{10}(|x/y|)$ as a function of $\alpha L$ for three specific cases, namely, (a) $\delta=0.0$, $\theta_3=0.0$ (b) $\delta=0.0$, $\theta_3=0.05$, (c) $\delta=0.001$, $\theta_3=0$ are shown in Figs.~\ref{fig2}(a),~\ref{fig2}(b) and \ref{fig2}(c), respectively. Curves from right to left are for $T_3=1.0,~0.49,~0.09$.
As can be seen from the solid curve in Fig.~\ref{fig2}(a) that the absence of the third cavity leads to the spectral singularity at $\alpha L=\sqrt{T_3}=1$. A decrease in $T_3$ leads to the same at lower values of $\alpha L$ and the singularity (being an inherent property of the $\mathcal{PT}$-symmetric component) survives. It is not surprising that a finite value of the detuning of the exterior cavity introduces the necessary dispersion to move the pole away from the real axis and the singularity is regularized (see Fig.~\ref{fig2}(b)). Recall that the true $\mathcal{PT}$-symmetry holds only for null detuning ($\delta=0$). A finite atomic detuning can kill the singularity even in a perfectly tuned exterior cavity (Fig.~\ref{fig2}(c)).
\subsection*{Effect of internal cavities in absence of the exterior cavity}
\begin{figure}[h]
\center{\includegraphics[width=8cm]{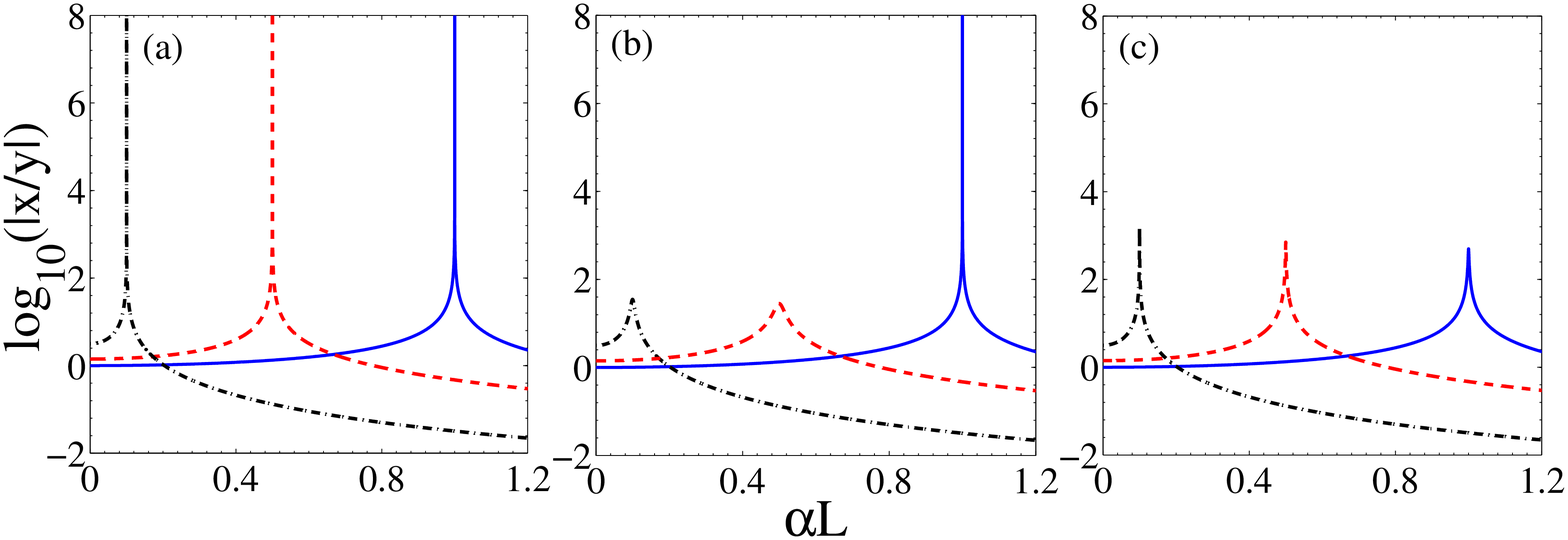}}
\caption{Effect of the internal cavities on the spectral singularity in absence of the external cavity ($T_3=1.0)$: (a) $\delta=0.0$, $\theta=0.0$ (b) $\delta=0.0$, $\theta=0.05$, (c) $\delta=0.001$, $\theta=0$. Curves from right to left are for $T=1.0,~0.5,~0.1$.} 
\label{fig3}
\end{figure}

Results for the transmission in absence of the exterior cavity ($T_3=1$) are shown in Fig.~\ref{fig3} for (a) $\delta=0,~\theta=0,$ (b) $\delta=0,~\theta=0.05$ and (c) $\delta=0.001,~\theta=0$. The curves from right to left in each panel are for $T=1.0,~0.5,~0.01$, respectively. As before, the perfectly tuned system displays the singularities at $\alpha L=T$ (Fig.~\ref{fig3}(a)).
A comparison of Fig.~\ref{fig2} and Fig.~\ref{fig3} reveals that the internal cavities play a more dominant role because of $T$ scaling of the location of the singularity as compared to the  $\sqrt{T}$ dependence in the previous case. It is also clear that atomic detuning plays the more significant role in limiting the infinite growth. A word of caution regarding the parameter values used in Fig.~\ref{fig2}(b) and Fig.~\ref{fig3}(b) would be in place since finite cavity detuning can be arbitrary for unity cavity transmission (cavity not present).
\begin{figure}
\center{\includegraphics[width=8cm]{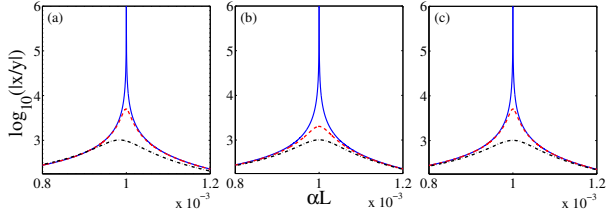}}
\caption{Removal of the spectral singularity in detuned systems for $T=T_3=0.01$ for (a) $\delta= 0.0,~ 0.01,~0.05$ with $\theta_3=\theta=0.0$, (b) $\theta_3= 0.0,~0.05,~0.1$ with $\theta=\delta=0.0$, and (c) $\theta= 0.0,~0.0001,~0.0005$ with $\theta_3=\delta=0.0$.}
\label{fig4}
\end{figure}
\begin{figure}[b]
\center{\includegraphics[width=6cm]{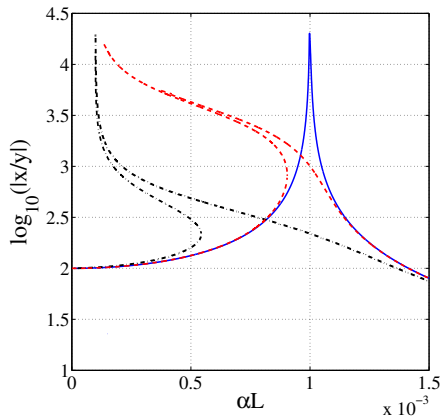}}
\caption{Regularization of the spectral singularity by nonlinearity in a cavity configuration ($T_3=T=0.01$,~ $\theta_3=0.005$ and $\delta=\theta=0$): $\log_{10}(|x/y|)$ as a function of $\alpha L$ for different values of $|y|$, namely, $|y|=10^{-6},~10^{-4}~,10^{-3}$. Curves from right to left are for increasing values of $|y|$.} 
\label{fig5}
\end{figure}
\subsection*{All three cavities present}
Consider the case when all the three cavities are present, say, with $T_3=T=0.01$. The results for this case are shown in Fig.~\ref{fig4} for different combinations of the cavity and atomic detuning parameters. Fig.~\ref{fig4}(a) shows how increasing atomic detuning can move the system away from the singularity with other detunings $\theta=\theta_3=0$. Similar behavior is displayed in the Figs.~\ref{fig4}(b) and \ref{fig4}(c) when $\theta_3$ and $\theta,$ are changed, respectively. It is evident from Figs.~\ref{fig4}(b) and \ref{fig4}(c) that the spectral singularity can be regularized with much lower values of $\theta$ than $\theta_3$. Fig.~\ref{fig4} thus demonstrates clearly the role of detuning parameters in regularizing the singularity in linear systems.
\begin{figure}[h]
\center{\includegraphics[width=8cm]{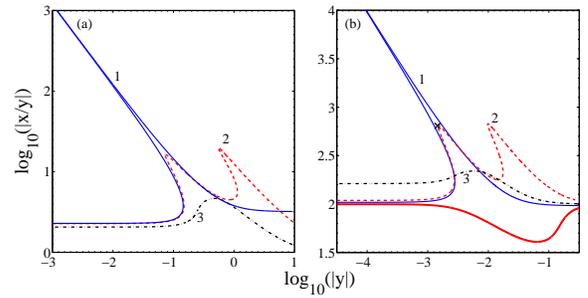}}
\caption{Regularization due to atomic detuning: $\log_{10}(|x/y|)$ as a function of $\log_{10}(|y|)$ (a) in a cavityless system ($T_3=T=1.0$) with $\alpha L=0.75$ for $\delta=0.0,~0.05,~0.2$ and (b) in presence of all the cavities $T_3=T=0.01$ with $\theta_3=\theta=0.0$ and $\alpha L=0.0002$ for $\delta=0.0,~0.15,~0.5$. Curves labeled $1-3$ are for the increasing values of $\delta$. In Fig.~\ref{fig6}(b) the point marked by a cross on curve 2 corresponds to very high transmission at low input, and the thick solid curve shows the diode action corresponding to curve $2$ in Fig.~\ref{fig6}(b).} 
\label{fig6}
\end{figure}
\begin{figure}[t]
\center{\includegraphics[width=8cm]{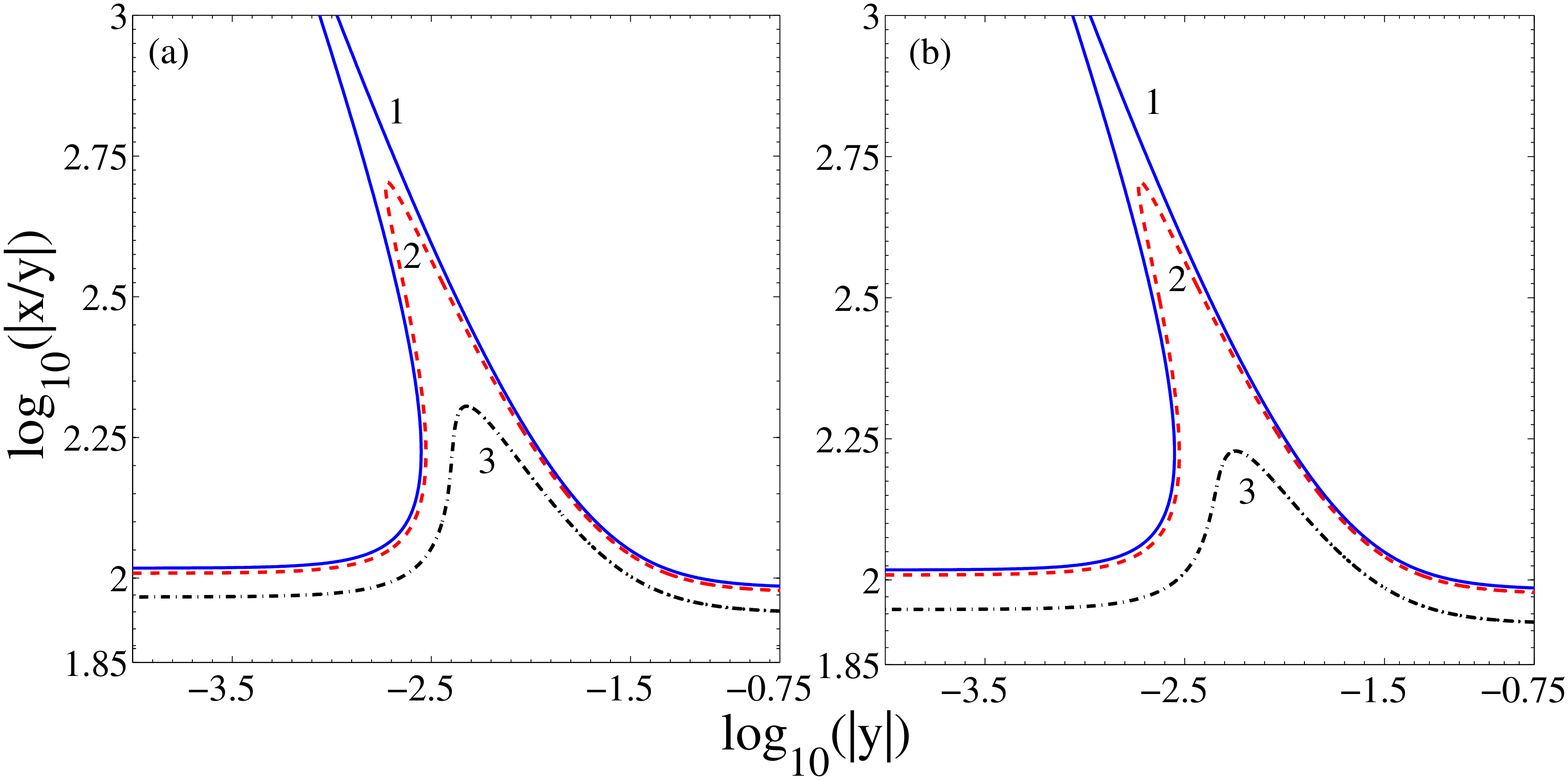}}
\caption{Effect of cavity detuning on spectral singularity in presence of all the cavities $T=T_3=0.01$ with $\delta=0.0,$ and $\alpha L=0.0002$ for (a) $\theta=0$, $\theta_3=0.0,~0.2,~0.5$ and (b)$\theta_3=0$, $\theta=0.0,~0.001,~0.003$. Curves labeled by $1-3$ are for increasing values of parameter of interest in each panel.} 
\label{fig7}
\end{figure}
\subsection*{Nonlinear response}
Till now we have analyzed the linear properties of the system when $|x|\rightarrow 0.$ We now look at the system response at larger power levels. As mentioned earlier, nonlinearity starts to play an important role and a saturation type nonlinearity as in our system can limit the infinite growth of the amplitudes. We show this regularization for larger values of $|y|$ in Fig.~\ref{fig5}. For a given input the input-output relation (Eq.~(\ref{eq1})) can yield multiple roots as in many other nonlinear systems \cite{sdgreview}. In order to avoid the infinite build-up near the singularity at low power levels, we consider a system slightly detuned from the spectral singularity and investigate the effects of nonlinearity on it. The system parameters were chosen as follows: $T_3=T=0.01$, $\theta_3=0.005$ with $\delta=\theta=0$. The transmission as a function $\alpha L$ for three different values of $|y|$ is shown in Fig.~\ref{fig5}. It is clear from Fig.~\ref{fig5} that nonlinearity can introduce dramatic changes to the response leading to different possible bistable scenarios. For example,  for $|y|=10^{-4}$ the resonance is bent towards left, while further increase in $|y|$ leads to an upward twist of the resonance. Effectively one now has the peak in transmission at a much lower value of $\alpha L$ at ($\alpha L=9.942\times 10^{-5}$). Thus cavity-designable nonlinear response can be handy for novel kind of switching phenomena in optical circuitry.
\par
The nonlinear response of the system is extremely sensitive to the detunings used. We have plotted transmission as a function incident field  (both in $\log_{10}$ scale) in Fig.~\ref{fig6}(a) and  Fig.~\ref{fig6}(b), where the two panels correspond to the case with no ($T=T_3=1$) and all cavities ($T=T_3=0.01$). In each panel we have presented the results for three distinct values of $\delta$. It is clear from figures that the ring cavity configuration can lead to bistable response for much lower values of $\alpha L$ ($=2.0\times10^{-4}$) as compared to $\alpha L$ (=0.75) for the cavityless situation. Moreover, there can be a multi-stable response (see curve 2) for suitable choice of $\delta$. The regularization feature for increasing values of $\delta$ can be easily seen from both the panels.%
\par%
Analogous response is also seen in transmission when cavity detuning is used as a parameter. This is shown in Fig.~\ref{fig7}, albeit with no multi-stable response. Fig.~\ref{fig7}(a) (Fig.~\ref{fig7}(b)) shows the transmission for three distinct detunings of the external (internal) cavity. As shown in Fig.~\ref{fig7} the diverging transmission can be regularized with much lower detuning values of the internal cavities than the external cavity. Note that this is consistent with the linear results shown in Figs.~\ref{fig4}(b) and \ref{fig4}(c). The control over the location and the height of the bent peaks in Fig.~\ref{fig6} and Fig.~\ref{fig7} can offer a true handle on the optical switching action and its contrast.
\par
Last but not the least, there have been ample discussions on nonreciprocity in linear and nonlinear systems. Nonreciprocity in reflection is a well known fact, while that in transmission is missing in linear and also in some Kerr nonlinear systems \cite{zafar-nrecp,sdg-nrecp,np-nrecp}. Nonreciprocity in transmission leading to optical isolation (diode action) was reported recently \cite{liu}. We now show that analogous nonreciprocity is exhibited by our system when the light is incident from the right or from the left. In the context of Eq.~(\ref{eq1}) it means that now $\alpha_1>0,$ and $\alpha_2<0,$ or the loss and the gain media are interchanged. The response for such a case is shown in Fig.~\ref{fig6}(b) by the thick solid line. It is clear that, the response is completely different for illumination from the left and the right. Evidently this can find many applications in optical circuitry. Another interesting point suggested by the nonlinear response, of say, Fig.~\ref{fig6}(b) is the following. Very large transmission can be realized at low powers. For example, the point marked by a cross on curve 2 of Fig.~\ref{fig6}(b) refers to such a situation when $|x/y|=6.4\times 10^{3}$ and $|y|=1.4\times 10^{-3}$. Of course one can achieve such a situation only after crossing the threshold (at $|y|=2.8 \times 10^{-3}$). The presence of cavity lowers the threshold value in comparison to that of cavityless system, for example, for curve 2 in Fig.~\ref{fig6}, the threshold value in presence of cavity is $|y|=2.8 \times 10^{-3}$, whereas for cavityless system $|y|=0.144$.\par
In conclusion, we have investigated a balanced gain-loss ring cavity system to show the presence of a singularity, which can be efficiently controlled by the cavity parameters and the atom-laser detuning. We compare various cases with or without the cavities to demonstrate this control on both linear and nonlinear response. We identified the atomic and cavity detuning parameters to be capable of removing the singularities. The regularization is also shown to be possible by changing the incident laser power, leading to skewed resonances and, in general multi-stability. Further, we demonstrated how to use it for the purpose of a controlled switch and optical diode action. Our results may find interesting applications in photonics devices.
\par
One of the authors  (SDG) is thankful to Girish S Agarwal for fruitful discussions. KNR is thankful to Department of Science and Technology for an INSPIRE Fellowship.

\end{document}